\documentclass[aps,prb,10pt,twocolumn,floatfix,superscriptaddress,showpacs,numerical,footinbib]{revtex4-1}

\usepackage{graphicx}
\usepackage{amsmath}
\usepackage{amssymb}
\usepackage[T1]{fontenc}
\usepackage[utf8]{inputenc}
\usepackage{hyperref}
\usepackage[pdftex]{color}

\newcommand{\ket}[1]{| #1 \rangle}
\newcommand{\bra}[1]{\langle #1 |}

\begin{document}
\title{Interaction driven phases in the half-filled honeycomb lattice: an infinite density matrix renormalization group study}
\author{Johannes Motruk}
\author{Adolfo G. Grushin}
\affiliation{\mbox{Max-Planck-Institut f\"ur Physik komplexer Systeme, N\"othnitzer Str.\ 38, 01187 Dresden, Germany}}
\author{Fernando de Juan}
\affiliation{\mbox{Materials Science Division, Lawrence Berkeley National Laboratories, Berkeley, CA 94720, USA}}
\affiliation{\mbox{Department of Physics, University of California, Berkeley, CA 94720, USA}}
\author{Frank Pollmann}
\affiliation{\mbox{Max-Planck-Institut f\"ur Physik komplexer Systeme, N\"othnitzer Str.\ 38, 01187 Dresden, Germany}}
\date{\today}
\begin{abstract}
The emergence of the Haldane Chern insulator state due to strong short range repulsive interactions in the half-filled fermionic spinless honeycomb lattice
model has been proposed and challenged with different methods and yet it still remains controversial.
In this work we revisit the problem using the infinite density matrix renormalization group method and report numerical evidence supporting 
i) the absence of the Chern insulator state, 
ii) two previously unnoticed charge ordered phases
and iii) the existence and stability of all the non-topological competing orders that were found previously within mean field.
In addition, we discuss the nature of the corresponding phase transitions based on our numerical data.
Our work establishes the phase diagram of the half-filled honeycomb lattice model tilting the balance
towards the absence of a Chern insulator phase for this model.
\end{abstract}
\maketitle

\section{Introduction}
\begin{figure}[t]
 \includegraphics[width=\columnwidth]{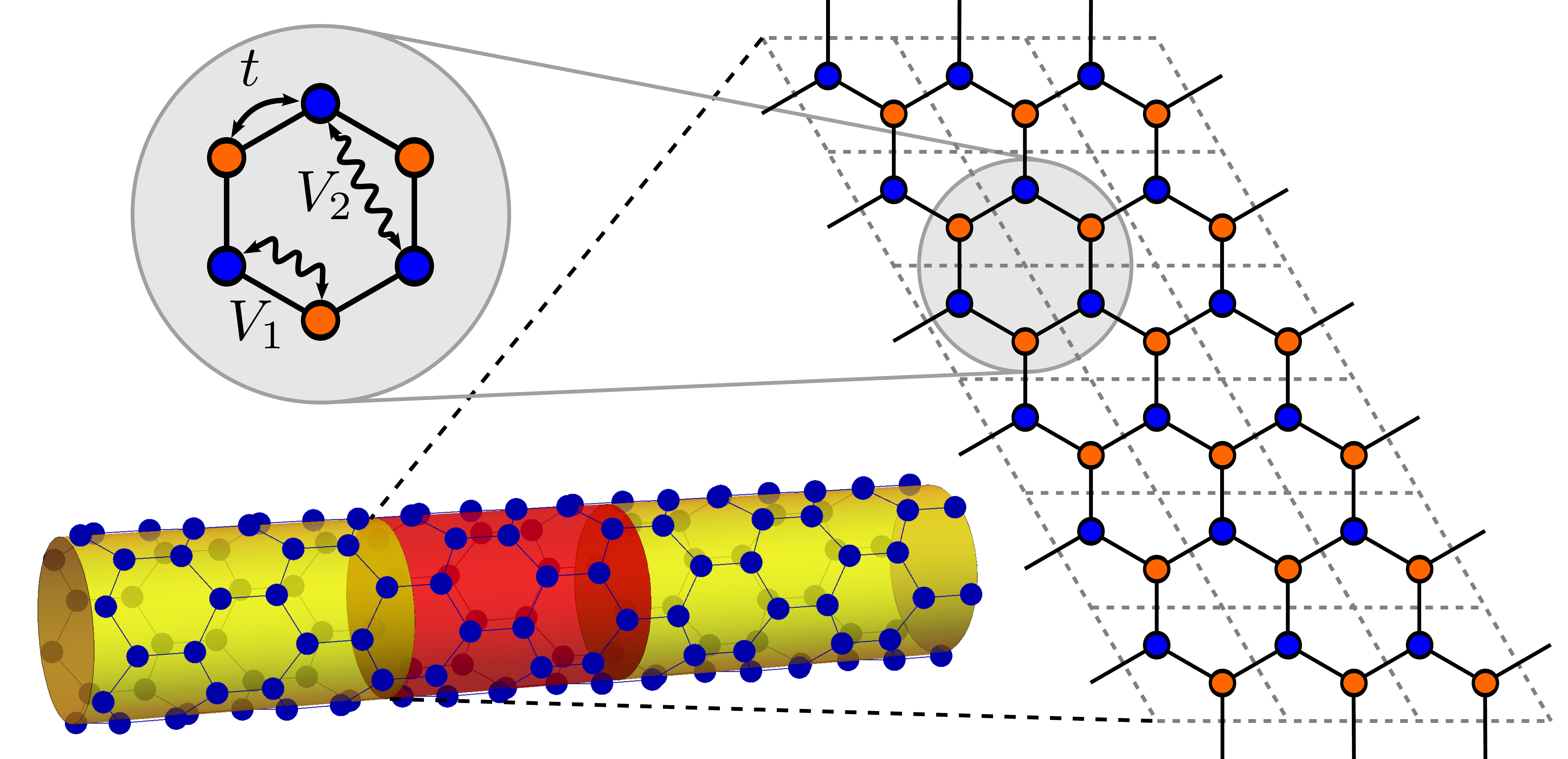}
 \caption{The top left illustration shows a schematic representation of the interacting tight binding model considered in this work:
fermions hopping in a half-filled honeycomb lattice with nearest-neighbor hopping $t$ (solid lines) and interacting via
 nearest- and next-to-nearest neighbor interactions, $V_{1}$ and $V_{2}$ respectively (curved lines) as defined by 
 \eqref{eq:H}.
 The right illustration shows the iDMRG unit cell used in this work with 
 $3 \times 6$ unit cells yielding a cylinder with a circumference of $L_{y}=12$ sites depicted in the bottom left illustration. 
 \label{fig:Defs}}
\end{figure}

\begin{figure}[b]
 \includegraphics[width=\columnwidth]{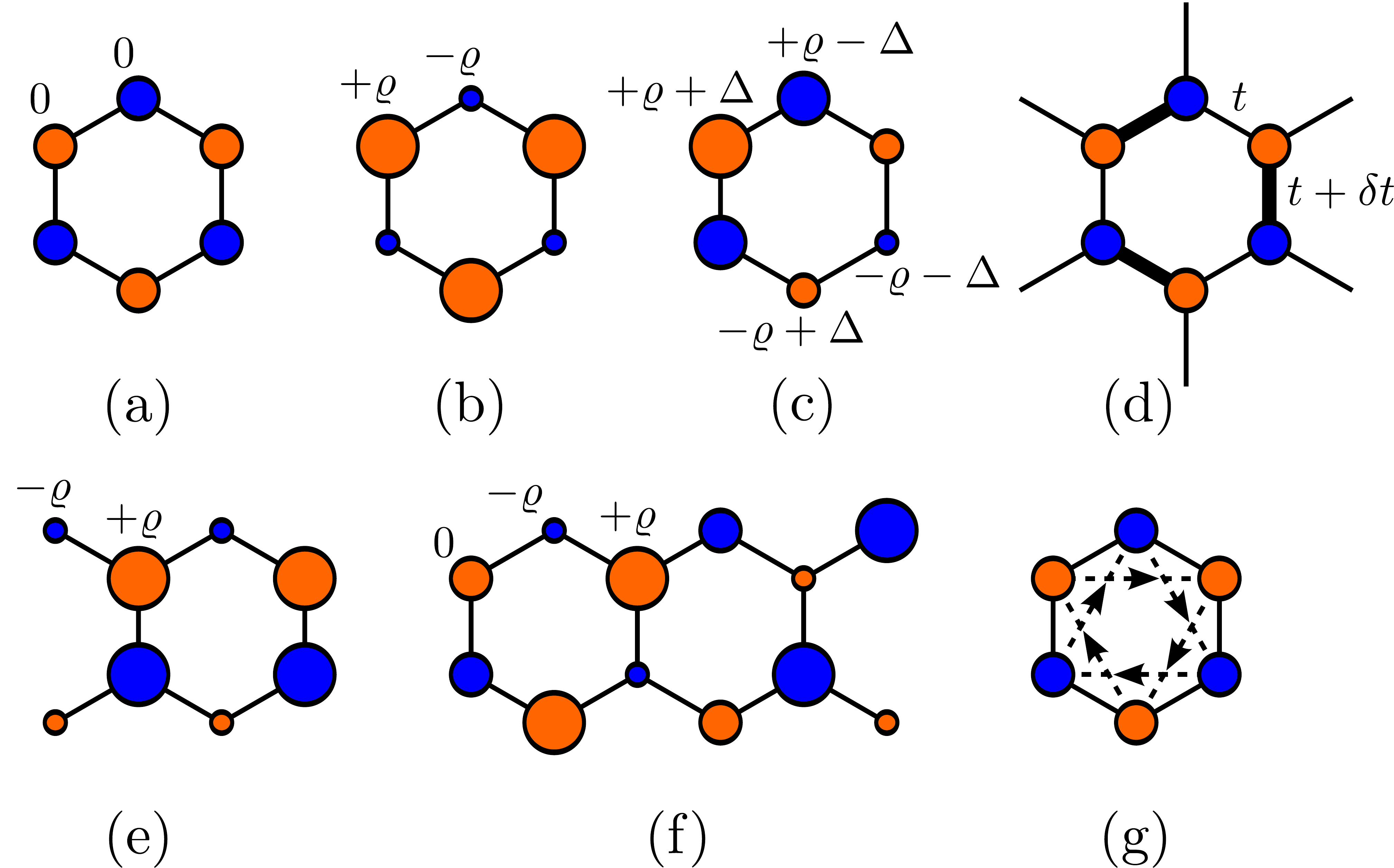}
 \caption{Different orders allowed in the half-filled spinless honeycomb lattice considered in this work:
 (a) semimetal, (b) charge density wave I (CDW~I), (c) sublattice charge modulated (CMs), (d) Kekul\'e, 
 (e) CDW~II, (f) CDW~III and (g) Haldane Chern insulator phases. 
 Phases (a)-(d) and (g) were found within mean field theory by Refs.~\onlinecite{RQHZ08,WF10,GCC13}.
The survival of the Haldane Chern insulator phase (g) was challenged within exact diagonalization with periodic boundary conditions
by Refs.~\onlinecite{GGNVC13,DH14} but found in Ref.~\onlinecite{DCH14} for open boundary conditions.
 The generic charge imbalance between the sublattices in phase (c) was not found within exact diagonalization.\cite{DH14}
 The two sublattices are depicted in blue and orange, $0 < \Delta < \varrho < 1/2$ with $\varrho + \Delta < 1/2$ describe the deviations from half-filling per site, i.e. $\langle n \rangle = 1/2 \pm \alpha$ with $\alpha \in \{\varrho, \varrho \pm \Delta \}$.
 \label{fig:orders}}
\end{figure}

Topological phases of matter are remarkably robust states; their responses to external fields are
governed by topological invariants and thus many of their most important properties are insensitive to local perturbations.\cite{HK10,QZ11}
Topologically protected responses result in intrinsically novel phenomena
including fractionalization of quantum numbers~\cite{Nayak2008} or transport governed by quantum anomalies~\cite{V03}
and can lead to diverse applications, ranging from fault tolerant quantum computation to spintronics.\cite{HK10,QZ11,Nayak2008}
Added to the remarkable experimental discoveries of new topological phases in both two and three dimensions,
these ideas have boosted a sustained and voluminous scientific effort in the last decade that attempts to classify them 
and determine when and how can they emerge.\cite{S14}\\
This work addresses a concrete question regarding the emergence of topological phases that has so far remained controversial: 
the existence of the Chern insulator phase triggered by short range repulsive interactions in the
fermionic spinless half-filled honeycomb tight binding model.
The Chern insulator state, first identified by Haldane~\cite{H88} in the particular case of the honeycomb lattice, is a zero field analogue of the
integer quantum Hall effect; the Hall conductivity contribution of each band is quantized in integer units of $e^2/h$ and determined by a topological invariant, the Chern number.
Realising a Chern insulator in nature is not a trivial task;\cite{CZF13} time-reversal must be broken with
the resulting magnetic flux averaging to zero over the unit cell, and hence over the entire sample.\\
In a proof of principle, a set of mean field studies~\cite{RQHZ08,WF10,PR12,GCC13} 
discussed how the Chern insulator can emerge from repulsive short range interactions. 
It was shown that spinless fermions hopping in the half-filled honeycomb lattice with nearest and next-to-nearest neighbor interactions, $V_{1}$ and $V_{2}$ respectively 
[see Fig.~\ref{fig:Defs}], displayed a Chern insulator phase as described by Haldane.\cite{H88}
The non-interacting realization of this Chern insulator phase has complex next-to-nearest neighbor hoppings [see Fig.~\ref{fig:orders}(g)].
Within the mean field paradigm it occurs in a region with $V_{2}>V_{1}$ where $V_{2}$ spontaneously breaks time reversal symmetry 
generating complex next-to-nearest neighbor hopping strengths.
The latter condition was shown not to be a generic requirement; Chern insulator phases can be realized 
in mean field with only $V_{1}$ at the expense of enlarging the unit cell and doping the system.\cite{CGV11,GCC13}
In addition, analogous topological phases have been obtained in other lattices such as the $\pi-$flux model~\cite{WF10,JGC13} and the Kagome lattice.\cite{WRW10}
\\

The important question that still remains to be answered is whether the interaction induced Chern state survives 
after the effect of quantum fluctuations is included.
The Haldane Chern insulator phase competes with more conventional but also interesting orders, 
that can jeopardize its emergence and are depicted schematically in Fig.~\ref{fig:orders}.
In the $V_{1} \gg V_{2}$ limit, a charge ordered state depicted in Fig.~\ref{fig:orders}(b) is expected, 
where the A and B sublattices are populated differently.\cite{RQHZ08,WCT14,LJY14b}
Being connected to the classical ground state at $V_{1}/t \to \infty$ with only one occupied sublattice, this state (CDW~I) has been found to be very robust both in mean field and exact diagonalization.\cite{GGNVC13,DH14}
For $V_{2}\gg V_{1}$ on the other hand it was shown under the mean field paradigm~\cite{GCC13} that the phase space originally attributed to the Haldane Chern insulator 
was severely reduced by the presence of a sublattice charge modulated state (or CMs) depicted in Fig.~\ref{fig:orders}(c).
The CMs phase was corroborated to survive quantum fluctuations in exact diagonalization~\cite{GGNVC13,DH14,DCH14} 
and within a variational Monte Carlo approach.\cite{DCH14}
The exact nature of the phase was further discussed in Ref.~\onlinecite{DH14} by computing the charge structure factor; 
its suppressed Fourier component at $\Gamma$ suggested the absence of charge imbalance between the sublattices and 
was therefore termed CM phase.
In addition, at intermediate $V_{1}\gtrsim V_{2}$ a Kekul\'e bond order~\cite{C00,HCM07,RH10,RJH13} 
depicted in Fig.~\ref{fig:orders}(d) was found to be stable within mean field theory.\cite{WF10}

This three-fold degenerate bond order triples the original honeycomb two atom unit cell breaking its translational symmetry.
Although numerical evidence consistent with such bond order was found in exact diagonalization,\cite{GGNVC13} further insights are needed to corroborate its existence in the thermodynamic limit.
Finally, the Haldane Chern insulator phase of Fig.~\ref{fig:orders}(g), stable within mean field, was found however to be absent in exact diagonalization with periodic boundary conditions~\cite{GGNVC13,DH14} and cluster perturbation theory~\cite{DH14} suggesting that quantum fluctuations indeed can destabilize its emergence.
Nonetheless, this interpretation was challenged by Ref.~\onlinecite{DCH14} that observed hints of this phase in exact diagonalization with open boundary conditions and variational Monte Carlo.\\

The contradictory numerical evidence regarding the existence of the Chern insulator phase in particular, and other competing phases in general, needs of an alternative approach
that includes quantum fluctuations while minimizing finite size effects.
In this work we therefore revisit the controversies left unsolved by previous studies using the infinite density matrix renormalization group method (iDMRG)~\cite{M08,W92,KZM13}. 
This variational method determines the ground state of systems of size $L_{x} \times L_{y}$ where $L_{x}$ is in the thermodynamic limit and $L_{y}$ goes beyond
what is achievable in exact diagonalization.
Traditionally a method for finding the ground state of one-dimensional systems,
iDMRG has recently been successfully applied to two-dimensional systems.
The infinite and finite DMRG method were indeed shown to be able to characterize the properties of fractional quantum Hall,\cite{ZMP13} 
$\mathbb{Z}_{2}$ quantum spin liquid,\cite{HSC14} chiral spin liquid \cite{HSC14b} and bosonic and fermionic fractional Chern insulating states.\cite{JWB12,CV13,ZKB13,GMZ15}
As long as entanglement remains low and the state has short correlation lengths the ground state
can be represented faithfully by a product of matrices --termed matrix product state (MPS)-- of a computationally affordable
dimension $\chi$, known as the bond dimension.\cite{M08,W92,KZM13}
This requirement is commonly met by gapped systems and in particular by the orders argued above to be expected instabilities in the half-filled 
honeycomb lattice of interest here.
Moreover, the iDMRG method deals with an infinite system and thus, given a suitable size of the unit cell used for the simulations, it can directly probe ground states that spontaneously break the original symmetries of the hamiltonian and the phase transitions among them, while allowing for quantum fluctuations to play a role. 
\\

Motivated by these advantages we have mapped out the $\left\lbrace V_{1},V_{2}\right\rbrace$ phase diagram for the half-filled honeycomb lattice in the infinite cylinder geometry
with the iDMRG method [see Fig.~\ref{fig:Defs}].
Our main findings are summarized next.
First we show compelling numerical evidence that supports the absence of the Chern insulator state in a wide region of $\left\lbrace V_{1},V_{2}\right\rbrace$ phase space.
Second, we find two new phases not reported previously neither within mean field~\cite{RQHZ08,WF10,GCC13} 
nor the first exact diagonalization results~\cite{GGNVC13,DH14,DCH14} and analyze their semiclassical features.
Third, we provide numerical support for the existence and stability of all the competing orders that were found previously within mean field.
In particular we characterize the two more controversial states, the CMs and Kekul\'{e} states, by computing bond and charge expectation values
and present further arguments of the existence of the former from the semiclassical large interaction limit.
Fourth, we provide numerical evidence to assess the first or second order character of the relevant phase transitions.
Concretely we address this issue by identifying the features that the correlation length $\xi$ and the entanglement entropy $S$ present as a function of $\left\lbrace V_{1},V_{2}\right\rbrace$.\\

This work is structured as follows. 
After describing the method and model in section~\ref{sec:modandmeth} we
characterize the different phases that we find in section~\ref{sec:phasediagram}.
The corresponding phase transitions among them are analyzed in section~\ref{sec:phasetransitions}
and we end with a discussion and prospect of our results in section~\ref{sec:discconc}.
Appendix \ref{sec:appendix} includes details of our semiclassical analysis as well a discussion of the CMs state order parameter.

\section{\label{sec:modandmeth}Model and Method}
We investigate a system of spinless fermions hopping on a honeycomb lattice with real nearest neighbor hopping $t\ge0$ interacting via nearest and next-to-nearest neighbor interactions 
$\left\lbrace V_{1},V_{2}\right\rbrace \ge \lbrace 0,0 \rbrace  $ respectively [see Fig.~\ref{fig:Defs}]. 
The Hamiltonian for this system can be written as
\begin{equation}
 H=-t\sum_{\left\langle i,j\right\rangle}(c^{\dagger}_{i}c^{\vphantom{\dagger}}_{j}+ \mathrm{h.c.})+
V_{1}\sum_{\left\langle i,j\right\rangle }n_{i}n_{j}+
V_{2}\sum_{\left\langle \left\langle i,j\right\rangle \right\rangle }n_{i}n_{j},
\label{eq:H}
\end{equation}
where $c_{i}^{\vphantom{\dagger}}$ $(c^{\dagger}_{i})$  annihilates (creates) an electron at the $i$-th site of the honeycomb lattice.\\
In order to find the ground state of the system in the $\left\lbrace V_{1},V_{2}\right\rbrace$ phase space
we employ the iDMRG algorithm~\cite{M08,W92,KZM13} on an infinite cylinder geometry [see Fig.~\ref{fig:Defs}].
As discussed above, such infinite geometry allows to probe ground states with spontaneous 
symmetry breaking while taking into account quantum fluctuations.
Feasible system sizes for our purposes are cylinders of circumference $L_{y} = 6,8,10$ and $12$, 
given the exponential growth of computational cost with the circumference.
For our calculations we choose $L_y=12$ and a unit cell of 36 sites depicted in Fig.~\ref{fig:Defs}.

We pick this geometry for the following two reasons.
First, we have to keep in mind the structure of the reciprocal space.
Since we work in the thermodynamic limit in the $x$-direction along the cylinder, the momentum in $x$-direction is a continuous quantity.
However in the finite $y$-direction around the cylinder, the momentum is a discrete variable.
In the non-interacting case, the model forms a Dirac semimetal and the Fermi surface is located at the 
$\mathbf{K}$ and $\mathbf{K}^{\prime}$ points.\cite{CastroNeto2009}
In order to capture the low energy physics correctly, it is of key importance that $\mathbf{K}$ and $\mathbf{K}^{\prime}$ are allowed momenta in our unit cell which implies that only $L_y=6$ and $12$ are suitable circumferences of our cylinder.
Second, we have the freedom of fixing the length of our unit cell in the $x$-direction since the computational for that only scales linearly.
By choosing three rings of 12 sites each, all orders are commensurate with our unit cell and a further enlargement would not lead to different results in the parameter region we investigated.
All data we show in the remainder of the text is computed for $L_y = 12$ and bond dimension $\chi = 1600$ unless otherwise stated.

\section{\label{sec:phasediagram}Phase diagram}
\begin{figure*}
 \includegraphics[width=\textwidth]{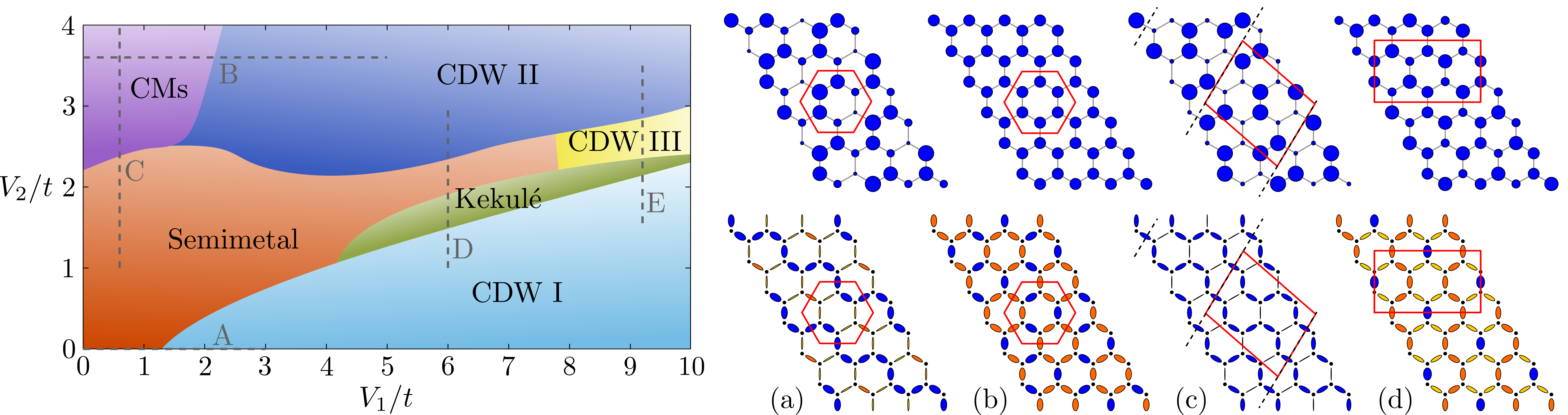}
 \caption{Left: Phase diagram obtained with iDMRG calculations on an infinite cylinder of circumference $L_{y}=12, \, \chi = 1600$.
 The phase boundaries, especially at second order transitions (see Sec.~\ref{sec:phasetransitions}) are to be taken with some error which can be estimated from Figs.~\ref{fig:cut_V2_0} to \ref{fig:cut_V1_9.2}.
 We draw sharp lines here for better clarity.
 Right: Representative charge and bond strength patterns for the four phases with largest unit cell discussed in the main text. 
 The area of the blue circles is proportional to the particle number expectation value on the respective site given by Eq. \eqref{eq:charge}. 
 The thickness of the ellipsoids on the bonds is proportional to the amplitude $t_{ij}$ between nearest neighbors defined by Eq.~\eqref{eq:bond}.
 The unit cells for each phase are depicted by the red polygons. 
 These correspond to (a) CMs phase with $V_1/t = 0.8, V_2/t = 3.2 $, (b) Kekul\'e phase with $V_1/t = 5.6, V_2/t = 1.6 $, (c) CDW~II phase with $V_1/t = 5.6, V_2/t = 3.2$,  
 (d) CDW~III phase with $V_1/t = 9.2, V_2/t = 2.5$. 
 For (c) dashed lines indicate defect lines of rotated hexagons that have zero energetic cost in the classical limit (see main text).
 \label{fig:phase diagram}}
\end{figure*}

We have mapped the phase diagram as a function of $\left\lbrace V_{1},V_{2}\right\rbrace$ with the method described above.
Our main results are summarized in Fig.~\ref{fig:phase diagram}.
We find six different phases: two new charge order phases labeled CDW~II and CDW~III and four previously 
reported mean field orders, the Kekul\'{e}, CMs, CDW~I and semimetal phases.\\
We have characterized each phase through their charge and bond ground state
expectation values.
At each site the charge expectation value is defined by 
\begin{eqnarray}
\label{eq:charge}
n_{i}=\left\langle c^{\dagger}_{i}c_{i}\right\rangle.
\end{eqnarray}
The bond ground state expectation value on the other hand is defined as
\begin{eqnarray}
\label{eq:bond}
t_{ij}=\left\langle c^{\dagger}_{i}c_{j}+\mathrm{h.c.}\right\rangle.
\end{eqnarray}
Next we discuss the features of the different phases in the phase diagram in Fig.~\ref{fig:phase diagram}.

\subsection{Semimetal phase \label{subsec:SM}}
We start by discussing the semimetal phase in the phase diagram shown in Fig.~\ref{fig:phase diagram}.
At $V_{1}/t=V_{2}/t=0$ the honeycomb lattice with nearest neighbor hopping its 
known to be described by a low energy theory in terms of two massless Dirac fermions.\cite{CastroNeto2009}
Short range interactions are irrelevant in the renormalization group sense~\cite{S94,KUP12} 
and therefore they can only drive a transition to an ordered state when they have a magnitude comparable to the nearest neighbor hopping strength $t$.
Such perturbative analysis guarantees that the semimetal is stable within
the region $\left\lbrace V_{1},V_{2}\right\rbrace \lesssim t$, only allowing for a uniform renormalization 
of the hopping strength $t$ by interactions.\\

For the semimetal phase in Fig.~\ref{fig:phase diagram}, and to numerical accuracy, we find by computing the charge expectation value
\eqref{eq:charge} that $n_{i}=1/2$ for all sites, indicating that this phase is not charge ordered.\\
From \eqref{eq:bond} and choosing $i,j$ to be nearest neighbors we find
a small asymmetry between bonds pointing along and around the cylinder axis,
with a relative difference of $\sim10^{-2}$.
Such asymmetry should --up to a very small effect due to the cylinder geometry which is always present for finite $L_y$-- vanish in a perfect semimetallic phase and indeed it is severely reduced as the bond dimension $\chi$ is increased.
This suggests that the cylinder geometry implemented in iDMRG artificially differentiates bonds in its two perpendicular directions.
The asymmetry induced by the finite bond dimension vanishes as $\chi$ is increased and thus it is not a physical effect.
This is consistent with a semimetal state; the logarithmic divergence of entanglement of a metallic state requires a matrix product state with $\chi\to\infty$. 
The bond asymmetry reduces the entanglement by shifting the Dirac cones away from the $\mathbf{K}$ and $\mathbf{K'}$ points.\cite{ACJ15}\\
Together, the previous numerical evidence are consistent with the semimetallic state.
We note that numerically the semimetal state extends beyond $\left\lbrace V_{1},V_{2}\right\rbrace \lesssim t$ towards
higher interaction strengths through a narrow semimetal trench in the phase diagram.
The larger size of this region at $L_{y}=6$ (not shown) suggest that it is likely to shrink as the circumference of the cylinder is increased but
a definitive statement requires going beyond the numerically accessible sizes. 

\subsection{Charge density wave I (CDW~I)}
Upon increasing $V_{1}$ we identify a charge density wave state labeled CDW~I in the phase diagram of Fig.~\ref{fig:phase diagram}.
This state has a two site unit cell and is characterized by a finite order parameter of symmetry $B_2$ under the symmetry group $C_{6v}''$ \cite{Basko08}
\begin{equation}
\label{eq:CDW}
B_2=\left\langle n_{A} \right\rangle-\left\langle n_{B}\right\rangle,
\end{equation}
that can be calculated using \eqref{eq:charge}, where $n_{A}$ and $n_{B}$ are the fermion densities in the $A$ and $B$ sublattice sites of the two site unit cell respectively. 
The resulting charge order is depicted schematically in Fig.~\ref{fig:orders}(b). 
For instance, at $V_{1}/t = 4$ and $V_{2}/t = 0.4$ we find that $B_2=0.817$.
The magnitude of $B_2$ increases with $V_{1}$ and to numerical accuracy this phase has no appreciable bond order.\\
For large $V_{1}\gg t$ such a state is a natural instability since the energy is minimized by a charge imbalance between the two sublattices.
Indeed, it has been found in a mean field approximation,\cite{RQHZ08,WF10,GCC13} exact diagonalization,\cite{GGNVC13,DH14,DCH14} 
and quantum Monte Carlo simulations.\cite{WCT14,LJY14b}

\subsection{Sublattice charge modulated phase (CMs)}
For small $t \ll V_{2}$ the ground state is classically degenerate. 
Within mean field it was shown that as long as $V_{2}>V_{1}$ the system chooses a charge ordered pattern with charge modulation 
of wavevector $\mathbf{K}$ or $\mathbf{K}'$, termed the CMs phase~\cite{GCC13} and depicted in Fig.~\ref{fig:orders}(c). 
The order parameters for any such modulation can be taken as the corresponding Fourier components of the charge expectation values of Eq.~\eqref{eq:charge},
namely $\left\langle n_{A,B}(\mathbf{K}) \right\rangle = \left\langle n_{A,B}(\mathbf{K}') \right\rangle^*$, where A and B indicate the two sublattices.
It is convenient to arrange these order parameters in a basis that has well defined transformation properties under the symmetry of the lattice.
Under the action of the symmetry group $C_{6v}''$, these order parameters transform as the four dimensional $G'$ representation \cite{Basko08,deJuan13}
\begin{align}
G_{1x}' = {\rm Re}[n_{A}(\mathbf{K}) -n_{B}(\mathbf{K})], \\
G_{1y}' =  {\rm Im}[n_{A}(\mathbf{K}) -n_{B}(\mathbf{K})], \\
G_{2x}' =  {\rm Im}[n_{A}(\mathbf{K}) +n_{B}(\mathbf{K})], \\
G_{2y}' =  {\rm Re}[n_{A}(\mathbf{K}) +n_{B}(\mathbf{K})].
\end{align}
The CMs state corresponds to a finite expectation value of both the $B_2$ order parameter and $G' = (1,0,0,0)$ and the states generated from these by operations of the symmetry group. 
A scalar order parameter for the CMs phase can be defined as $B_2 {\rm Re}[(G_1')^3-3G_1'(G_2')^2]$, where $G_i' = G_{ix}'+iG_{iy}'$ (see appendix A).
The CMs structure is schematically shown in Fig.~\ref{fig:orders}(c). 
It minimizes the large cost attributed to $V_{2}$ by reversing two nearest neighbor dimers at the expense of paying the small energetic cost determined by $V_{1}$.
Within exact diagonalization it has been argued that the CMs state survives quantum fluctuations.\cite{GGNVC13,DH14,DCH14}
However, the sublattice charge imbalance predicted in mean field~\cite{GCC13} was suggested to vanish in Ref.~\onlinecite{DH14} 
due to the absence of an enhanced Fourier component of the charge structure factor at the $\Gamma$ point.
We note that the CMs scalar order parameter defined above quantifies the charge imbalance since $B_{2}$ has to be non-zero for it not to vanish.\\
With the iDMRG method we find that the CMs appears directly above the semimetallic phase~[see Fig.~\ref{fig:phase diagram}].
A sample of the six site unit cell charge and bond pattern obtained from \eqref{eq:charge} for $V_1/t = 0.8, V_2/t = 3.2 $ is shown in Fig.~\ref{fig:phase diagram}(a).
The circles at each site have an area proportional to the strength of the charge at the given site, while the thickness of the links between the sites represent the bond strengths.
This bond and charge pattern survives in the region labeled CMs of the phase diagram and coincides with that obtained within mean field theory~\cite{GCC13} depicted schematically in Fig.~\ref{fig:orders}(c).
The corresponding charge values for $V_1/t = 0.8, V_2/t = 3.2 $ are given by $\varrho=0.364$ and $\Delta = 0.092$.
Moreover, we find a clear sublattice imbalance that grows as $V_{2}$ is increased that justifies the label CMs rather than the simpler CM.
We note also that the bond ordering of this phase is associated to the charge order: two neighboring sites with similar high (or low) charge densities
that deviate from half-filling suppress the hopping between them due to the large (small) number of filled (empty) states at that site. 
\\
We have also accounted for the existence of the CMs state from a strong coupling perturbation theory analysis. 
By exactly diagonalizing the interaction part of $H$ in Eq.~\eqref{eq:H} and then finding the first order hopping corrections $t/V_{1,2}$ in perturbation theory, 
we determined the ground state $\left.|GS\right>$ of a $3\times3$ cluster. 
We then computed different scalar correlation functions of the charge order parameters, and found that only $\left<GS|B_2 {\rm Re}[(G_1')^3-3G_1'(G_2')^2]|GS\right>$ is finite. 
This confirms that in the strong coupling limit, the ground state is indeed of the CMs form. The details of the procedure are presented in appendix \ref{sec:appendix}. 

\subsection{Kekul\'{e} bond order}
The next phase we identify is the Kekul\'{e}
bond order.
Like the CMs, it has a six site unit cell depicted schematically in Fig.~\ref{fig:orders}(d)
with uniform charge order and two types of bonds, strong and weak.
Under the mean field approximation the state arises between the CMs and the CDW~I state.
Although in previous exact diagonalization hints of this state are also observed,
the evidence supporting its occurrence is not entirely conclusive~\cite{GGNVC13} and
its tripled unit cell turns the analysis of larger clusters challenging.

Within iDMRG we find that this state is stable in a finite but smaller region compared to both 
exact diagonalization and mean field.
In contrast to the anisotropy found in the semimetal phase, the Kekul\'e order remains stable as the bond dimension $\chi$ is increased and stays finite upon extrapolation to infinite $\chi$.
We show a representative of the state's numerically obtained charge and bond order patterns in Fig.~\ref{fig:phase diagram} (b) 
calculated with Eqs.~\eqref{eq:charge} and \eqref{eq:bond}, respectively.
The bond thickness is proportional to its strength.
From Fig.~\ref{fig:phase diagram}(b) it is apparent that the state has indeed two bond strengths arranged as in Fig.~\ref{fig:orders}(d) 
and no charge order.
Therefore the system chooses to break the original two site unit cell translational symmetry 
falling into one of the three distinct Kekul\'{e} ground states.\cite{WF10}
At $V_{1}/t=5.6$ and $V_{2}/t=1.6$ the two types bonds are found numerically to be 
$t_{1}=0.522$ and $t_{2}=0.428$.

\subsection{Charge density wave II (CDW~II)}
All phases that we have described so far do not differ
from those predicted by mean field~\cite{RQHZ08,WF10,PR12,GCC13} 
and the first exact diagonalization studies.\cite{GGNVC13,DH14,DCH14}
At finite $V_{1}$ and sufficiently large $V_{2}$ we observe
a different charge order state (CDW~II), see Fig.~\ref{fig:phase diagram} left panel.\\
To gain a first insight on this phase before analyzing the numerical data it is instructive to discuss its classical limit.
A strong coupling analysis (see appendix \ref{sec:appendix} for details) reveals quite generically 
that at $t=0$ there are two classical regimes separated by the $V_{1}=4V_{2}$ line. 
For $V_{1}>4V_{2}$ the state CDW~I is favored with the same charge pattern as in Fig.~\ref{fig:orders}(b) but classical occupations 0 or 1.
At $V_{1}=4V_{2}$ there exists a classically degenerate ground state manifold, to be discussed in the next section in the context of the CDW~III phase.
The analysis for $V_{1}<4V_{2}$ with $t=0$ reveals that there are essentially two types of degenerate states according to their translational symmetry.
The first is a stripe-like phase with a four atom unit cell depicted in~Fig.~\ref{fig:orders}(e) (see also~\onlinecite{CL15}) that is six fold
degenerate in the thermodynamic limit.
The second is a set of states that can be understood by taking the stripe-like phase and rotating $60^{\circ}$ hexagons along a defect line that should wrap
around the chosen cluster.
Classically, the energy per site of these two types of states does not have first order hopping corrections and it is given by
$E_{\mathrm{CDW~II}} = V_2/2+V_1/4 + \mathcal{O}(t^2/V_{1,2})$ (see appendix \ref{sec:appendix} for details). 
The first nontrivial order in perturbation theory depends on the chosen cluster; eventually 
the degeneracy between the six-fold degenerate stripe-like phase and the phase with defects will be lifted, 
making one of them more favorable at finite $t$.\cite{CL15}
The question then becomes which type of ground state does the system choose in the thermodynamic limit.\\

Although we cannot attempt to fully answer this question some light can be shed through our iDMRG numerical data. 
We indeed find that both configurations, with and without defects are ground states with very similar energy,
even close to the transition to the semimetal phase.
A sample of one of this charge ordered patterns is shown in Fig.~\ref{fig:phase diagram}(c) for $V_{1}/t=5.6$ and $V_{2}/t=3.2$.
This particular sample state has one defect line that wraps twice around the cluster (indicated by dashed lines); rotating the hexagons along these
lines by $60^{\circ}$ in an anticlockwise direction we recover the more symmetric stripe-like configuration in Fig.~\ref{fig:orders}(e).
As with the CMs state, we find that its bond order follows directly from its charge order, suppressed for neighboring sites that have
equal filling.\\
By initializing the algorithm with each inequivalent classical ground state configuration it is possible to calculate to high precision each 
energy to then find, by comparison, the lowest energy state at that particular $\left\lbrace V_{1},V_{2}\right\rbrace$.
We have determined that within the CDW~II region the lowest energy is often achieved by a superposition of two classical ground states
and their particle-hole conjugates.
As expected from general arguments~\cite{KZM13} this state shows a degenerate low energy entanglement spectrum.
Inspection of the two point density-density correlation functions indicates that 
the more favored superposition of classical ground states are those which favor a uniform charge density pattern that at the same time
distinguishes the bonds around and along the cylinder geometry.

\subsection{Charge density wave III (CDW~III)}
At high $V_{1}$ and between the previously discussed CDW~II and CDW~I states
we find a third type of charge order with a twelve site unit cell labeled CDW~III.
A representative pattern of its charge and bond order as obtained numerically 
with \eqref{eq:charge} and \eqref{eq:bond} respectively
is shown in Fig.~\ref{fig:phase diagram}(d).
As schematically represented in Fig.~\ref{fig:orders}(f), this state has three 
different occupations: half-filling and $1/2\pm\varrho$.
For the particular case of $V_{1}/t=9.2$ and $V_{2}/t=2.5$ we find that $\varrho=0.223$.
Its bond order is determined by the charge occupations 
similar to the CMs and CDW~II phases.
As with the CDW~II phase, this phase escaped
identification in the original mean field~\cite{RQHZ08,WF10,GCC13} 
and the first subsequent exact diagonalization~\cite{GGNVC13,DH14,DCH14} studies 
due to its large unit cell.\\
The emergence of the CDW~III can be understood from a semiclassical point of view (see appendix A).
At $t=0$, the classical ground state of lowest energy for a cluster commensurate with this phase is either in the CDW~II or CDW~I class, 
and the two phases are separated by the line $V_{1}=4V_{2}$. 
Exactly at this line, however, a third classical state is degenerate in energy with these but, unlike the previous two, is affected by first order quantum corrections. 
This implies that around the line $V_{1}\sim4V_{2}$ there is a finite strip of a new phase in the phase diagram once finite $t$ is included. 
We have checked that the two point correlation function calculated numerically from the iDMRG data and semiclassically 
agree qualitatively, resulting in the charge distribution of the CDW~III shown in Fig.~\ref{fig:phase diagram}(d).
We find furthermore that these are consistent with those reported in the latest exact diagonalization study.\cite{CL15} 

\section{\label{sec:phasetransitions} Phase transitions}
In order to unravel the character of the corresponding phase transitions we now study two quantities that are sensitive
to a phase change, the entanglement entropy $S$ and the correlation length $\xi$.
The entanglement entropy is defined by
\begin{equation}
 S = -\mathrm{Tr} \rho_L \ln \rho_L,
\end{equation}
where $\rho_L = \mathrm{Tr}_R \ket{\psi}\bra{\psi}$ is the reduced density matrix of the left semi-infinite half $L$ of the cylinder after tracing out the right half $R$.
The correlation length $\xi$ can be computed from the resulting state of the iDMRG calculation according to Ref.~\onlinecite{KZM13}.
Both quantities are expected to show a finite discontinuity when crossing a first order transition while the correlation length diverges with increasing bond dimension at a second order critical point.\cite{KZM13}
In what follows, we focus on the lines labelled by cuts A--E in the phase diagram in Fig.~\ref{fig:phase diagram}.

\subsection{Cut A: semimetal--CDW~I}

\begin{figure}
 \includegraphics[width=\columnwidth]{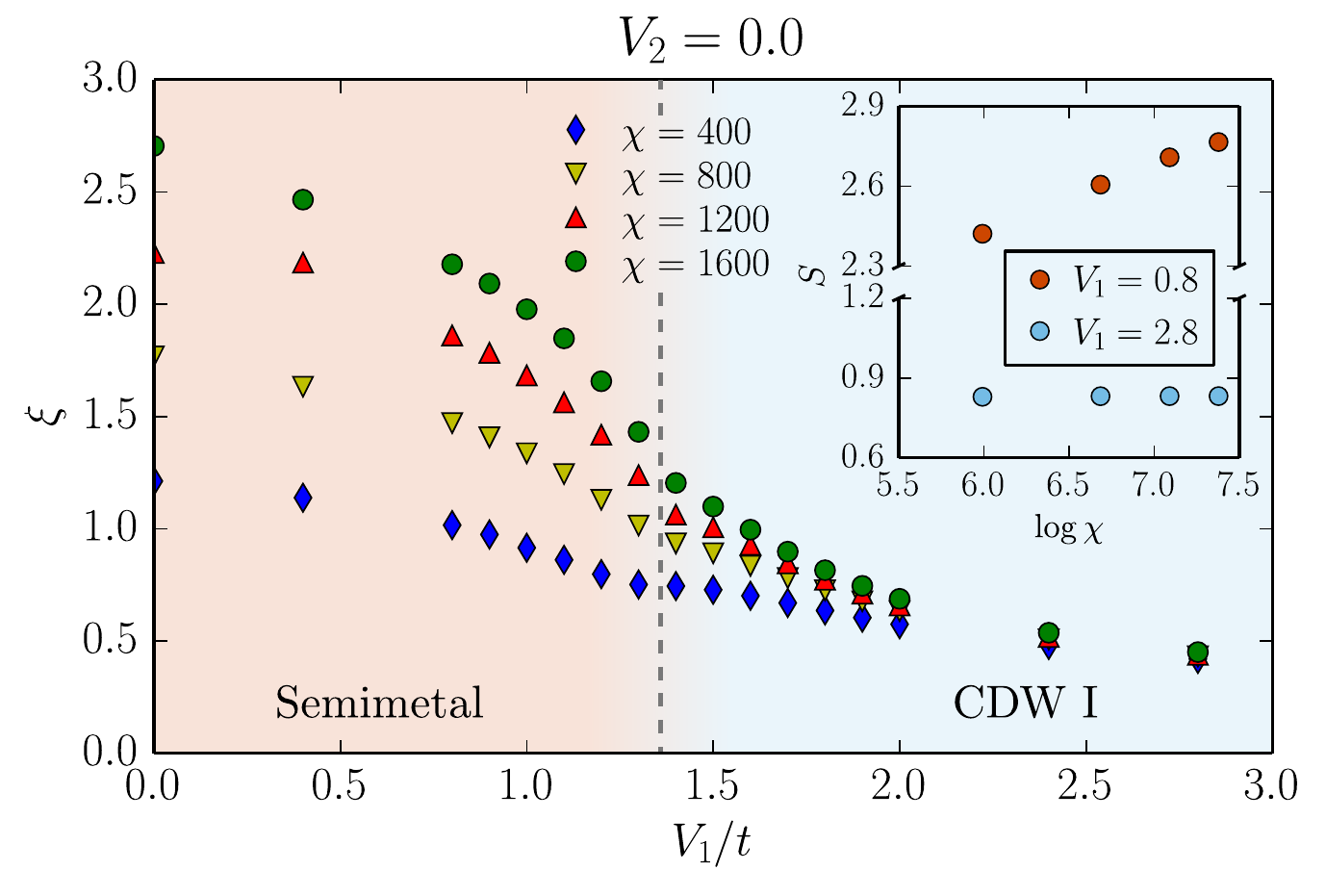}
 \caption{Correlation length at $V_2=0$, labeled cut A in Fig.~\ref{fig:phase diagram} probing the transition between 
 the semimetal and CDW~I phases. 
 The smooth behavior of $\xi$ indicates a second order transition. 
 The dashed grey line at $V_1/t = 1.356$ shows where the transition has been detected in quantum Monte Carlo simulations.\cite{WCT14,LJY14b}
 Inset: Scaling of the entanglement entropy $S$ with the bond dimension $\chi$ for two values of $V_{1}$ in the semimetal and CDW~I phase. 
 Unlike the CDW~I phase, the semimetal phase presents the typical critical entanglement scaling $S\propto \ln\chi$.
 \label{fig:cut_V2_0} }
\end{figure}

The first horizontal cut, labeled A addresses the character of the phase transition between the semimetallic
phase and the simplest charge density wave (CDW~I) by fixing $V_{2}=0$.
This phase transition has been previously addressed with the quantum Monte-Carlo method~\cite{WCT14,LJY14a,LJY14b} 
and was determined to be of second order character.
The transition point with divergent correlation length was determined to be at $V_{1}/t = 1.356$ via finite size scaling.
In Fig.~\ref{fig:cut_V2_0} we show the correlation length as a function of $V_{1}$ 
for different values of the bond dimension $\chi$.

Firstly, for $V_{1} \lesssim 1.5t$ we observe that the correlation length drops as a function of $V_{1}$
and diverges as the bond dimension $\chi$ is increased.
This behaviour is expected for a critical state such as the semimetal phase;  
the logarithmic divergence of entanglement of a metallic state requires an 
matrix product state with $\chi\to\infty$.
For $V_{1} \gtrsim 1.5t$ the correlation length continues to drop but has no longer a significant 
dependence on $\chi$.
This is characteristic of a gapped phase such as the charge density wave.
%
%
In the inset of Fig.~\ref{fig:cut_V2_0}, we plot the entanglement entropy as a function of the logarithm of the bond dimension for two values of $V_1$ representative for the two phases.
In the semimetal phase for $V_1=0.8$, we observe a finite entanglement scaling of $S \propto \ln \chi$ characteristic of a critical phase,\cite{FES2009} whereas the entanglement entropy at $V_1=2.8$ in the gapped 
CDW I phase is independent of the bond dimension.

The crossover between the two phases is smooth, signaling a second order phase transition, in agreement with 
quantum Monte Carlo studies.\cite{WCT14,LJY14b}
With iDMRG it is however not possible to pin point the exact value of $V_{1}$ 
where the transition happens via finite size scaling due to the few cylinder sizes 
we have available, as discussed in section \ref{sec:modandmeth}.
However, from Fig.~\ref{fig:cut_V2_0} it is possible to define a crossover region of $1.3 \lesssim V_{1}\lesssim 1.5$, 
consistent with the quantum Monte Carlo data, \cite{WCT14,LJY14b} where the transition occurs.

\subsection{Cut B: CMs--CDW~II}

\begin{figure}[t]
 \includegraphics[width=\columnwidth]{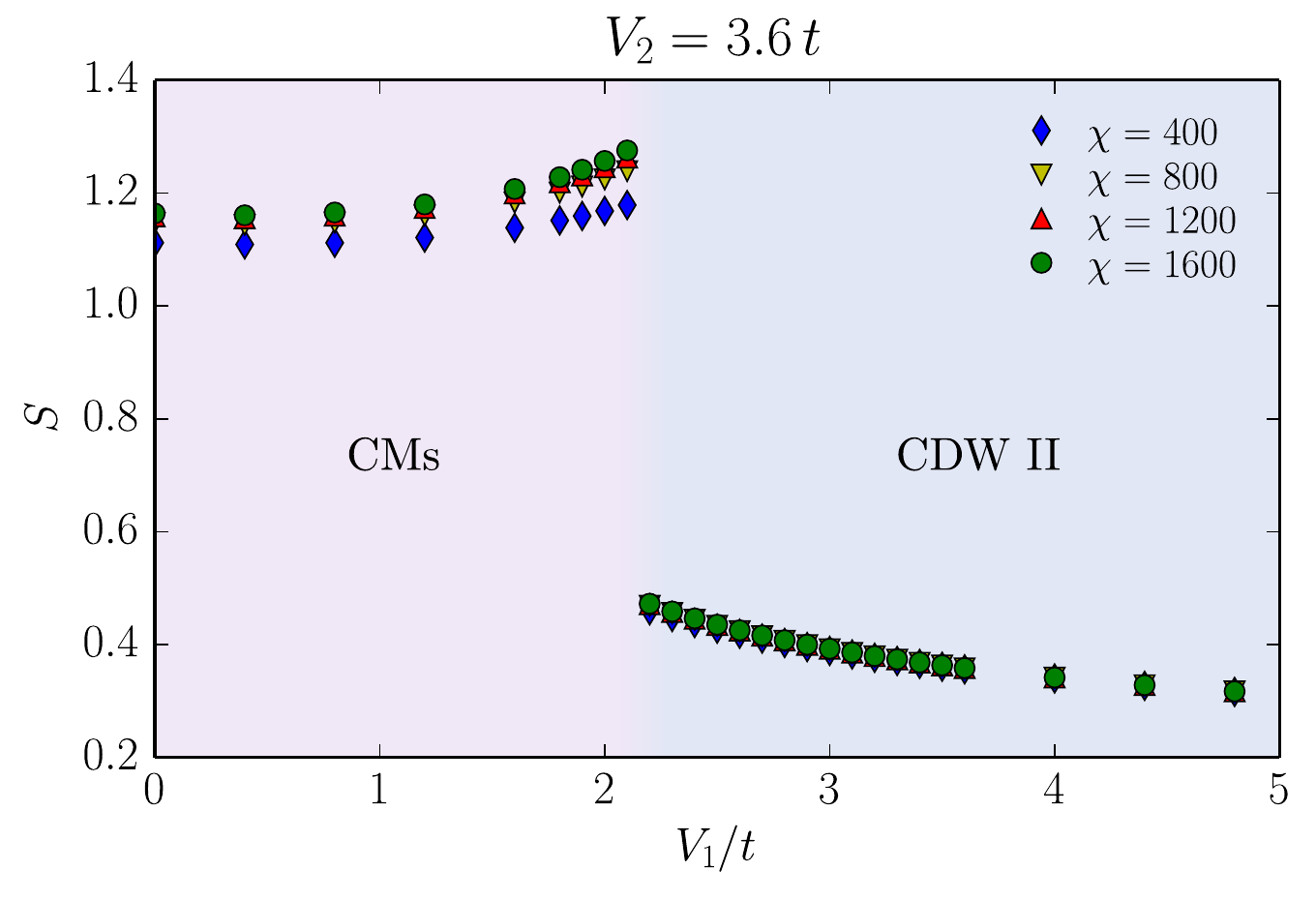}
 \caption{Entanglement entropy at $V_2/t=3.6$, labeled cut B in Fig.~\ref{fig:phase diagram}. The finite discontinuity of $S$ at $V_{2}/t\approx 2.2$ clearly signals a first oder transition between the CMs and CDW~II phases. \label{fig:cut_V2_3.6}}
\end{figure}

In cut B we fix $V_{2}=3.6t$ and study the phase transition between
the CMs and the CDW~II phases.
The entanglement entropy is shown in Fig.~\ref{fig:cut_V2_3.6} 
as a function of $V_{1}$.
This quantity shows a clear discontinuity at $V_{2}/t\approx 2.2$ signaling a direct first order
phase transition between these two phases.
We observe the same behavior when looking at the correlation length not presented here.
Additionally, the energy of the ground state as a function of $V_2$ shows a kink at the same critical $V_2$ which further supports the presence of a level crossing at that point.
As expected for gapped phases, the entanglement entropy only depends very weakly on the bond dimension, in the CDW~II phase even a very low $\chi$ of 400 is sufficient to faithfully represent the state.

A first order phase transition is also expected from the strong coupling approach.
As detailed in appendix \ref{sec:appendix}, the energy per site of the CM phase in a $3\times3$ cluster is 
$E_{\text{CMs}} = V_2/2+V_1/3-0.248t$ for $V_1/V_2<3/2$, while the energy per site of the CDW~II phase is $E_{\text{CDW~II}} = V_2/2+V_1/4$.
At $t=0$, $V_1=0$ both states have the same energy.
At finite $t$, the CM phase has lower energy due to the first order quantum correction, which is absent for CDW~II. 
Since $E_{\text{CMs}}$ grows faster with $V_1$ than $E_{\text{CDW~II}}$, there must be a crossing at a critical value
of the interaction signaling a first order phase transition into the CDW~II state.

\subsection{Cut C: Semimetal--CMs}
\begin{figure}
 \includegraphics[width=\columnwidth]{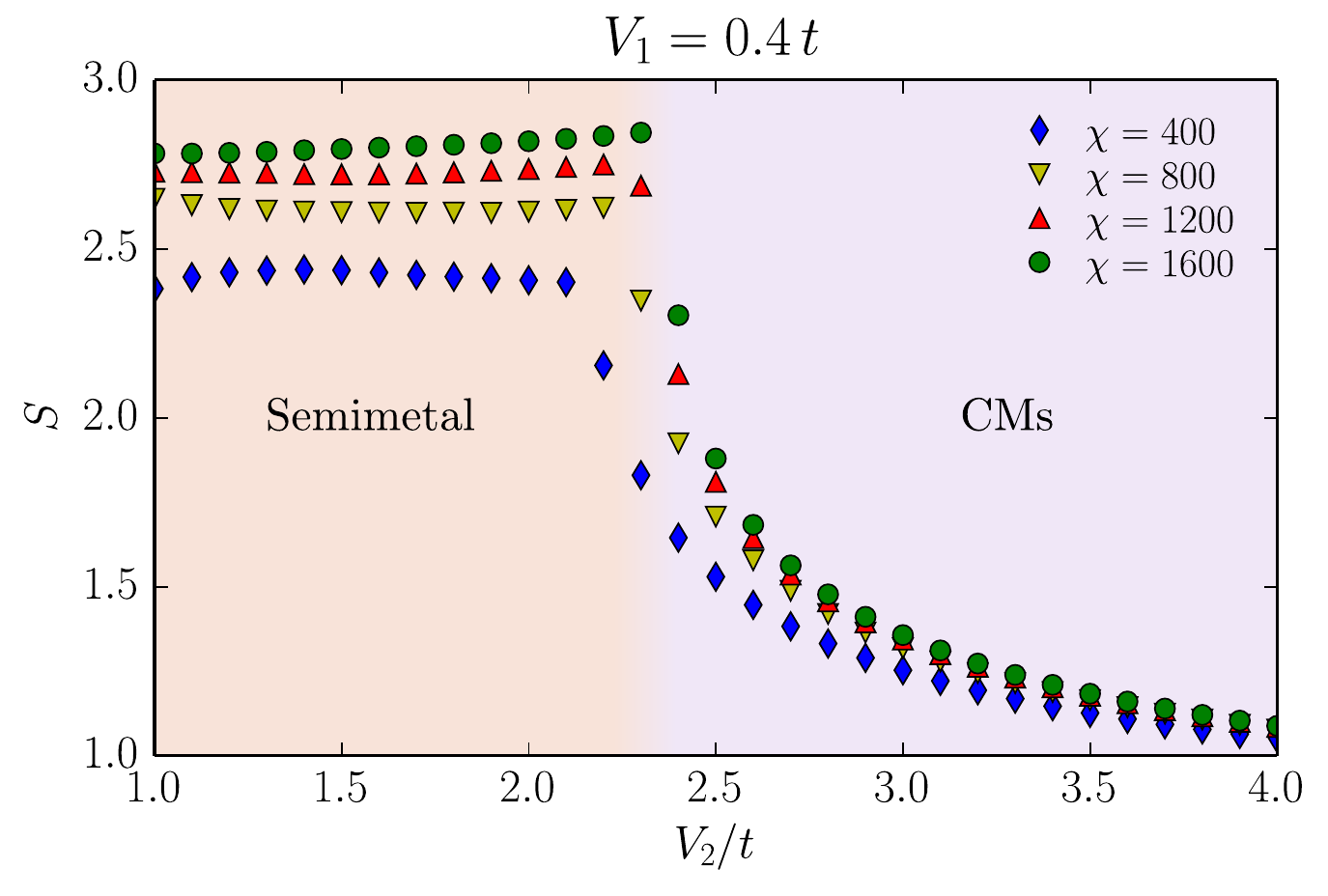}
 \caption{Entanglement entropy at $V_1/t=0.4$, labeled cut C in Fig.~\ref{fig:phase diagram}, probing the phase transition between the semimetal and CMs phases. 
 Going towards the phase boundary from high $V_2$, the entanglement entropy smoothly approaches the value in the semimetal indicating a second order transition. \label{fig:cut_V1_0.4}}
\end{figure}

The cut at $V_{1}/t=0.4$, labeled cut C in Fig.~\ref{fig:phase diagram}, 
probes the transition between the semimetal phase and the CMs phase.
In Fig.~\ref{fig:cut_V1_0.4} we present the entanglement entropy $S$ as a function of $V_{2}$, the interaction that drives the phase transition.

As explained in Sec.~\ref{subsec:SM}, the entanglement entropy of 
the semimetal phase depends strongly on the bond dimension $\chi$.
At $V_{2}/t\approx 2.3 $ we observe a sharp transition
to a decaying entanglement entropy that does not depend strongly on $\chi$
as $V_{2}$ is increased.
However, the change in entanglement entropy is not as abrupt as in the CMs--CDW~II case of the previous subsection as would be expected from a level crossing in a first order transition.
We can see a kink in $S$, but for decreasing $V_2$ it smoothly approaches the value of the semimetal phase.
In addition, the energy is a smooth function $V_2$ which together with the previously mentioned behaviour of $S$ suggests that the transition is of second order.
The correlation length not depicted here shows a similar smooth behavior as in the transition between 
the semimetal and CDW~I phases in Fig.~\ref{fig:cut_V2_0} which provides further evidence in favor of a second order transition.

\subsection{Cut D: CDW~I--Kekul\'{e}--Semimetal--CDW~II}

\begin{figure}
 \includegraphics[width=\columnwidth]{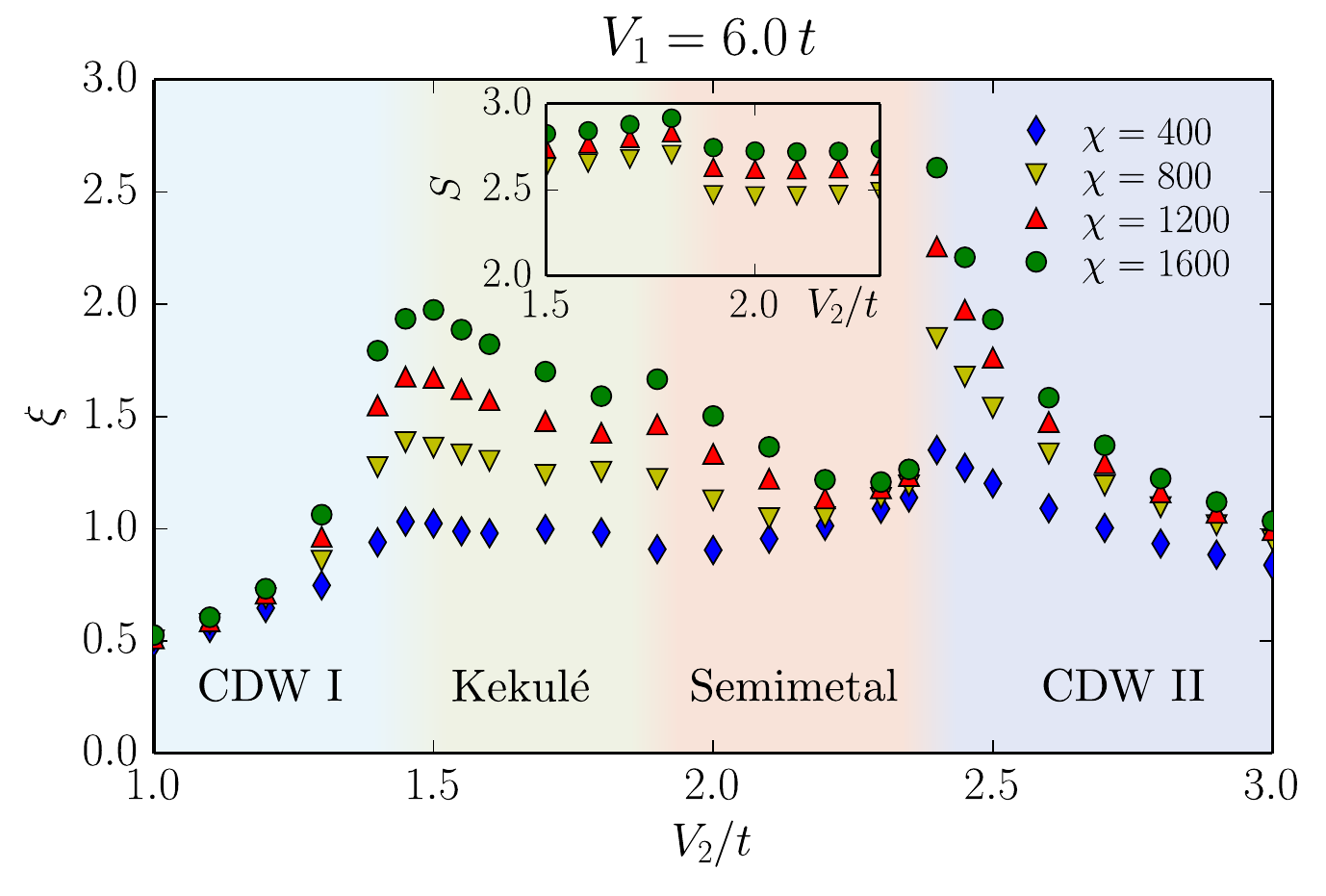}
 \caption{Correlation length at $V_1/t=6$, labeled cut D in Fig.~\ref{fig:phase diagram}. 
 In this plot, several phase transitions are made visible. 
 With increasing $V_2$, the CDW~I phase turns into the Kekul\'e phase which itself goes into the semimetal phase.
 The transition from semimetal to CDW~II can be identified by the peak in the correlation length. 
 Inset:  The vicinity of the Kekul\'e--semimetal transition showing the discontinuity of the entanglement entropy as a function of $V_{2}$ \label{fig:cut_V1_6}}
\end{figure}

The next cut we focus on is labeled cut D in Fig.~\ref{fig:phase diagram} at $V_1/t = 6$. The correlation length as a function of $V_2$ is shown in Fig.~\ref{fig:cut_V1_6}.
Starting from low $V_2$ we first cross the phase boundary between the charge density wave CDW~I and the bond-ordered Kekul\'e phase.
As expected for a gapped phase, the correlation length deep in the CDW~I phase depends only weakly on the bond dimension.
As we approach the phase transition, this behaviour changes since the many body gap is closing.
The strongly increasing correlation length which peaks at $V_2/t \approx 1.5$ suggests a second order transition between the CDW~I and Kekul\'e phases.
Moreover, we can see the presence of a remnant charge order on top of the bond modulation in the Kekul\'e phase close to the critical point.
This signals a slow onset of charge ordering as a finite size effect as we approach the phase boundary from the Kekul\'e side which would be in contradiction to a level crossing for which the charge order should change suddenly.
A similar second order transition was reported in the related antiferromagnetic Heisenberg model between a N\'eel and a plaquette phase.\cite{GvdBN13,ASHCML11,CL15}
This transition was conjectured to show deconfined quantum critical behavior.\cite{SVBSF04,Yao}

The fact that the correlation has not fully converged with $\chi$ even in the center of the the Kekul\'e region indicates that this phase is strongly entangled.
This can be attributed to the fact that it is not a strong coupling phase in which the gap is determined by the interaction but rather by the hopping energy scale.
As $V_{2}$ is increased we cross the boundary between the Kekul\'e and the semimetal phase at $V_2/t \approx 1.9$.
From the data presented in Fig.~\ref{fig:cut_V1_6} it is not possible to reach a conclusion about the nature of the transition.
However, from the Landau theory perspective the semimetal--Kekul\'e transition has to be first order for the following reason.
The Kekul\'e bond order is in the twofold representation $E_1' = (E_{11}',E_{12}')$ \cite{Basko08} and the lowest order non-trivial term in the Landau free energy is the scalar $(E_{11}')^3-3E_{12}'(E_{11}')^2$ which is of third order, therefore signaling a first order transition. 
A first order transition is also consistent with the entanglement entropy which displays a small discontinuity when crossing the phase boundary (see inset of Fig.~\ref{fig:cut_V1_6}).

In the semimetal phase, we observe an astonishingly small correlation length for a critical state.
As explained in Sec.~\ref{subsec:SM} the hopping strengths around and along the cylinder are strongly renormalized.
This shifts the Dirac nodes away form the $\mathbf{K}$ and $\mathbf{K'}$ points thus leading to an effectively gapped state for a finite cylinder circumference with small correlation length.

Finally, at $V_2/t \approx 2.4$ the semimetal turns into the CDW~II.
The diverging correlation length on the CDW~II side of the transition points towards a second order phase transition.
Moreover, the strong dependence of $\xi$ on the bond dimension suggests that the ground state is becoming critical as we approach the transition point, 
especially compared to the behaviour of $\xi$ deep in the CDW~II phase.
On the other hand, from the semimetal side the correlation length is discontinuous, which could signal a first order transition.
However, the latter observation has to be taken with care since the semimetal phase, being a critical state, cannot be represented faithfully with a finite bond dimension.

\subsection{Cut E: CDW~I--Kekul\'e--CDW~III--CDW~II}
\begin{figure}
 \includegraphics[width=\columnwidth]{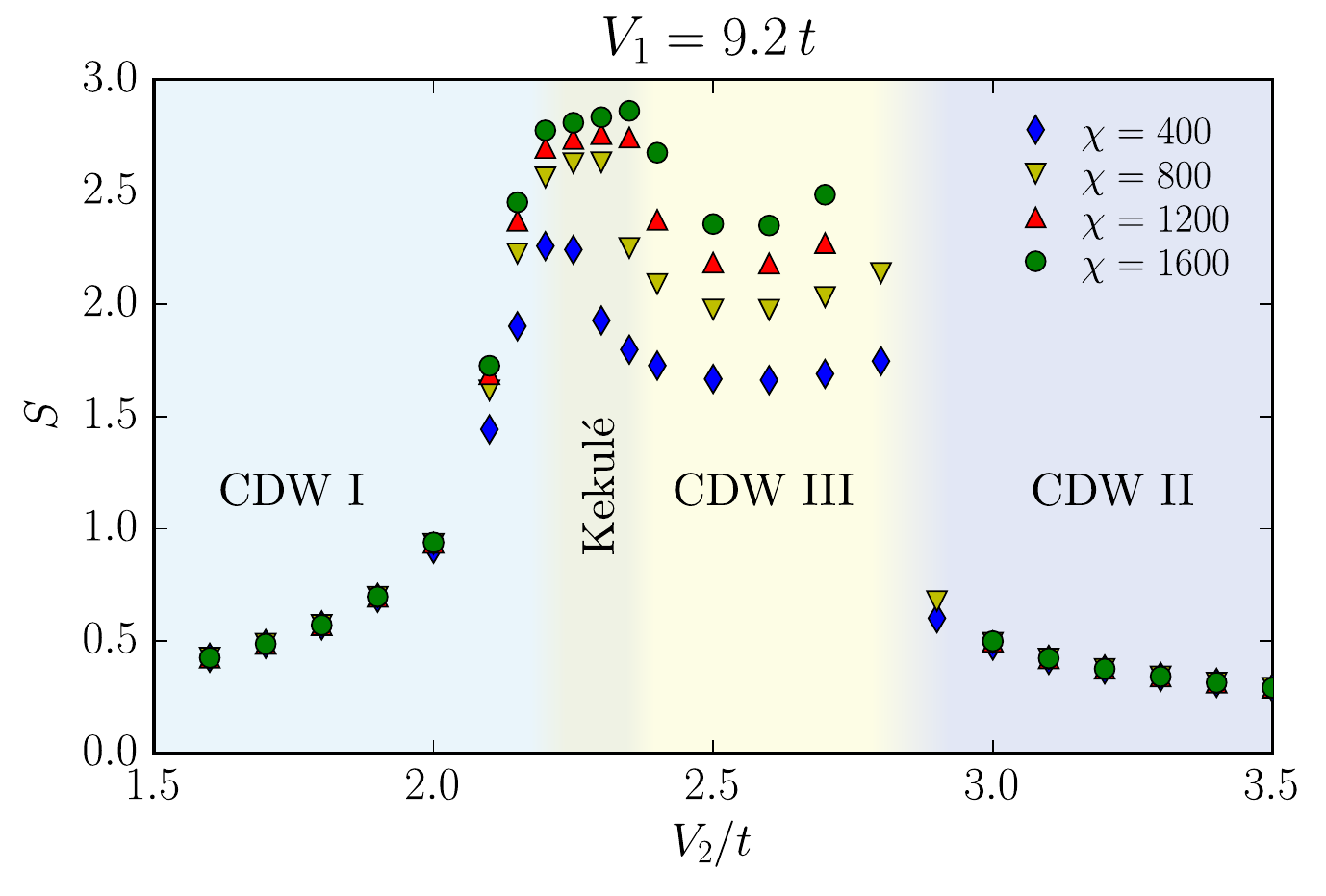}
 \caption{Entanglement entropy at $V_1/t=9.2$, labeled cut E in Fig.~\ref{fig:phase diagram}. 
 The smooth behavior of $S$ at the phase boundary between CDW~I and Kekul\'e phase indicates a second order transition in accordance with the data of the correlation length in 
 Fig.~\ref{fig:cut_V1_6}. 
 At $V_2/t \approx 2.8$, the entanglement entropy is discontinous, which clearly indicates a first order transition between the CDW~III and CDW~II phases. 
  \label{fig:cut_V1_9.2}}
\end{figure}
The last cut we address is labeled cut E in Fig.~\ref{fig:phase diagram} and fixes $V_1/t = 9.2$. 
The entanglement entropy is presented in Fig.~\ref{fig:cut_V1_9.2}. 
With increasing $V_2$ we cross the second order phase transition from the CDW~I to the Kekul\'e phase at $V_2/t \approx 2.2$ discussed above.
The entanglement data shown in Fig.~\ref{fig:cut_V1_9.2} supports this statement since $S$ smoothly increases in the CDW~I phase as we approach the phase boundary.

At $V_2/t \approx 2.4$ the system undergoes the transition to the CDW~III phase.
Unfortunately, a conclusive statement about the order of the critical point cannot be drawn from the present data; especially since the entanglement entropy in the CDW~III phase still shows a significant dependence on the bond dimension.

The last phase boundary in this cut lies between the CDW~III and CDW~II phases at $V_2/t \approx 2.8$.
Here, we can see a clear discontinuity of the entanglement entropy, similar to that in cut~B (Fig.~\ref{fig:cut_V2_3.6}).
Moreover, we see a kink in the energy when crossing the phase boundary.
Together these features indicate a first order phase transition.

\section{\label{sec:discconc}Discussion and Conclusions}

The spontaneous emergence of the Haldane Chern insulator phase~\cite{H88} due to interactions in the half-filled spineless honeycomb lattice
model has been questioned by different methods since its proposal.\cite{RQHZ08,WF10,GCC13,GGNVC13,DH14,DCH14}
In this work we have used the iDMRG method and have mapped out the phase diagram of this model.
The algorithm allows for quantum fluctuations to play a role but minimizes the finite size effects due to its intrinsic infinite cylinder geometry.

One of our main results is the absence of the Haldane Chern insulator state in this model.
Even when initializing the iDMRG calculations with a time reversal symmetry broken chiral wave function, the state did not remain stable and evolved into the respective 
competing phases upon applying the algorithm.
It seems therefore that quantum fluctuations indeed jeopardize emergence of the CI phase and work against
the naive mean field expectations.
Although there is no reason to believe that the situation is different for the $\pi$-flux model, which also hosts a
mean field Chern insulating phase that is absent in exact diagonalization,~\cite{WF10,JGC13}
there is more hope for other models with quadratic band point touchings.
In particular, the interacting kagome lattice hosts a Chern insulator state within mean field theory when the filling is tuned to
the quadratic band point touching between the flat band and one of the two dispersive bands.\cite{WRW10}
Its presence is predicted within the renormalization group approach~\cite{SF08,SYF09} which guarantees
its robustness at sufficiently low interaction strength.
Adequate substrate engineering can also lead to such quadratic band point touching, potentially favoring the Chern insulator state.\cite{MVB14}
\\
In addition, we have theoretically accounted for two novel competing phases, the CDW~II and CDW~III.
Their potentially large unit cells turns their identification within exact diagonalization or mean field studies 
challenging.
On the one hand the CDW~II phase stems from a degenerate strong coupling phase that has no first order correction in powers of $t/V_{1,2}$.
Although the degenerate manifold includes a distinguishable stripe-like phase, our iDMRG calculations find that a superposition state 
between classical ground states has lower energy.
The particular superposition that is realized seems to be determined by the cylinder geometry where bond strength along
and around the cylinder are inequivalent. 
The CDW~III on the other hand, occurs in a finite region around the classical transition between
the CDW~I and II states defined by the line $V_{1}=4V_{2}$, and has a clear twelve site unit cell.
In this region, the first order corrections $t/V_{1,2}$ lift the classical degeneracy and stabilize CDW~III state.
Both the semiclassical limit and our numerical iDMRG data show consistent charge structure factors confirming
the semiclassical nature of the state.
The CDW~III phase presents as well a distinctive bond ordering that was understood from its charge order.

From the evidence presented in this work we have established that the CMs state occurs from first
order perturbation theory in powers of $t/V_{2}$ of an otherwise classically degenerate ground state.
This part of our work establishes the existence and robustness 
of this single particle charge order beyond numerical approximations.
Our semiclassical analysis, corroborated by the iDMRG numerical evidence,
provides an explanation of why this phase has a many-body gap of the order of the hopping strength $t/V_{2}$,
as was previously noticed,~\cite{DH14,DCH14} rather than determined by $V_{2}$ .
Both the semiclassical treatment and the numerical data we obtain establish that the CMs state
has a finite sublattice imbalance, a feature that was still under dispute.\cite{GCC13,DH14}
Moreover, we have reported a characteristic bond order not addressed in previous studies.
As in the CDW~II and CDW~III cases such bond order is determined by the CMs charge order pattern.
Our results show that the CMs is stable for small $V_{1}$ and finite $t$.
Increasing the former or reducing the latter leads to the CDW~II ground state through a first order phase transition.\\
Furthermore, we have established the existence of the Kekul\'{e} bond order which occurs in a region of the phase 
diagram that appears to be smaller than that predicted by mean field and exact diagonalization.\\
Finally, we have used the fact that the iDMRG method treats an infinite system as opposed to a finite cluster to analyse 
the different phase transitions using in particular the entanglement entropy $S$ and correlation length $\xi$.
The conclusions drawn from analyzing these quantities have been complemented by the continuous 
or discontinuous dependence of the ground state energy on the $\left\lbrace V_{1},V_{2}\right\rbrace$ interactions.
However, conclusions about the order of phase transitions from and to the semimetal phase should be taken with care due to the gapless nature of the semimetal ground state. \\

To conclude we have established the phase diagram of spinless fermions hopping on the half-filled honeycomb lattice
with nearest- and next-to-nearest neighbor interactions and characterized the phase transitions among the different phases with iDMRG.
Our results provide solid evidence that Chern insulating phases are far more elusive than previously thought and
so alternative routes are necessary to drive these kind of topological states from strong electronic correlations in general.
\section{Acknowledgements}

We thank A. L\"auchli for discussions, useful insights regarding the nature of the CDW~II and CDW~III phases
and sharing consistent exact diagonalization results prior to publication.\cite{CL15}

\appendix

\section{\label{sec:appendix} Strong coupling perturbation theory}

In this appendix we discuss the exact diagonalization method to determine the ground state in the strong coupling limit $t \ll V_{1,2}$. 
Consider splitting the Hamiltonian in Eq. \eqref{eq:H} into $H =H_V + H_t$ with
\begin{align}
 H_t &=-t\sum_{\left\langle i,j\right\rangle}(c^{\dagger}_{i}c^{\vphantom{\dagger}}_{j}+ \mathrm{h.c.}) \\
H_V &= V_{1}\sum_{\left\langle i,j\right\rangle }n_{i}n_{j}+
V_{2}\sum_{\left\langle \left\langle i,j\right\rangle \right\rangle }n_{i}n_{j}. 
\label{eq:Hsplit}
\end{align}
In the strong coupling limit, we can obtain the ground state of $H$ by diagonalizing $H_V$ first and considering $H_t$ as a perturbation. 
The eigenstates of $H_V$
\begin{equation}
H_V \psi_{n,m} = E_n \psi_{n,m},
\end{equation}
where $m$ accounts for degeneracies, are simple to compute because charge is conserved at every site in the absence of hopping. 
$H_V$ is thus already diagonal in the occupation basis, i.e.
\begin{equation}
\left| \psi^{n,m}\right> = \prod_{C^{n,m}_i = 1} c^\dagger_i \left|0\right>,
\end{equation}
 with $C^{n,m}_i=0,1$ the occupation coefficients for the $n,m$ eigenstate. 
The classical ground state manifold is spanned by $\left|\psi^{0,m}\right>$, with $m=1,\dots,M$.   

For small $t/V_{1,2}$, we can disregard the classical states with $n>0$, and project the $H_t$ Hamiltonian into the classical ground state manifold. 
Dropping the label $n=0$ from now on
\begin{equation}
(\tilde{H}_t)_{m_1m_2} =  \left< \psi^{m_1} \right| H_t \left| \psi^{m_2} \right>.
\end{equation}
We can now diagonalize $\tilde{H}_t$ and select the eigenstates of lowest energy $\epsilon_0$
\begin{equation}
\tilde{H}_t v_\alpha = \epsilon_0 v_\alpha,
\end{equation}
where $\alpha = 1,\ldots,D$ and $D$ is the true ground state degeneracy. The final ground state of $H$ is spanned by
\begin{equation}
\left| GS\right>_\alpha = \sum_{m=1}^M v^m_\alpha \left| \psi^m\right>.
\end{equation}
Obtaining the coefficients $v^m_\alpha$ is relatively simple because the effective size of the Hilbert space $M$ is much smaller than the full size of the Hilbert space of $H$.
To distinguish different phases, we recall that in finite size exact diagonalization there is no spontaneous symmetry breaking, because all states related by symmetry are degenerate and will be present in the ground state manifold. 
In the simpler case when first order quantum corrections are zero, a phase can be characterized by inspection of the classical charge patterns of every eigenstate. 
With quantum corrections, however, the ground state can only be characterized by computing correlation functions evaluated in the ground state, from which order parameters can be obtained. 
Correlation functions for charge order parameters can be expressed in general as
\begin{equation}
\rho_{ij\ldots} = {\rm tr} \left< GS \right| c^\dagger_i c_ic^\dagger_j c_j \ldots \left| GS \right>, 
\end{equation}
which is given explicitly by
\begin{equation}
\rho_{ij\ldots} = \sum_{m,\alpha} (v_m^\alpha)^*v_m^\alpha C^m_i C^m_j \dots
\end{equation}
The behavior of correlation functions can then be used to identify the different phases. 
We have used this method to diagonalize three different clusters: two 12 site clusters with the periodicities of the CDW~II and CDW~III phases, 
and a cluster of $3\times3$ unit cells or 18 sites, i.e. the $L_y=6$ version of the cluster in Fig.~\ref{fig:Defs}. 
The results can be summarized as follows. \\

In the case of the CDW~II cluster, we find only two possible classical ground states with degeneracies 8 and 2 for generic values of $V_1/V_2$. 
The projection of $H_t$ in these subspaces is zero for both cases, which means that both are stable strong coupling phases.
The first can be identified with the classical version of the CDW~II state, where the two values of the occupancies are 0 or 1.
The 8 states correspond to the 8 equivalent ways to ensemble the CDW~II pattern in the given unit cell: two of them correspond to the stripe-like phase
shown schematically in Fig.~\ref{fig:phase diagram}(c) while the other six correspond to those states with defect lines determined by rotated hexagons, as described in sec.~\ref{sec:phasediagram} E. 
The second is the CDW~I state, where the two states have a single sublattice (A or B) fully occupied. The energy per site of these states is given by
\begin{align}
E_{\mathrm{CDW~II}} &= \frac{1}{2}V_2 + \frac{1}{4}V_1 & & \tfrac{V_1}{V_2} < 4 \label{cdwii} \\
E_{\mathrm{CDW~I}} &= \frac{3V_2}{2}   & & 4 < \tfrac{V_1}{V_2}\label{cdwi}
\end{align}
\\
In the case of the CDW~III cluster, for generic values of $V_1/V_2$ we find two possible classical ground states with degeneracies 2 and 2. 
The first correspond to the stripe-like phase mentioned before, which is also commensurate with this cluster, and is therefore assigned the label CDW~II. The second corresponds to CDW~I. The energies per site remain the same as those in Eqs.~\ref{cdwii}-\ref{cdwi}. 
In addition, in this cluster there is another classical state with degeneracy 18 and energy $E^*=\frac{5}{6}V_2 + \frac{1}{6}V_1$ which requires consideration.  
This energy is always higher than $E_{\mathrm{CDW~II}}$ or $E_{\mathrm{CDW~I}}$, except at the special point $V_1=4V_2$ where the three cross, $E_{\mathrm{CDW~II}}=E_{\mathrm{CDW~I}}=E^*$. 
The reason why this state must be considered is that quantum corrections split it to first order in $t$, thus lowering its energy. Since neither CDW~II or CDW~I is affected by $t$ to first order, there is a small region around $V_1=4V_2$ where the energy of this state is lowest. 
Inspection of the phase diagram in Fig.~\ref{fig:phase diagram} suggests that this is the CDW~III phase. The ground state of this phase has degeneracy one and energy
\begin{align}
E_{\mathrm{CDW~III}} &= \frac{5V_2}{6} + \frac{V_1}{6} - 0.236t   & & \tfrac{V_1}{V_2} \sim 4 \label{cdwiii}
\end{align}
To confirm the nature of this state, we have computed the Fourier transform of the correlation function $C(q) = \sum_{ij} e^{i(x_i-x_j)q} \rho_{ij}$, and confirmed that it compares favorably with the Fourier transform of the charge pattern of CDW~III shown in Fig.~\ref{fig:phase diagram}(d).\\
In the case of the $3\times3$ cluster, at $t=0$ we find four possible classical ground states as a function of increasing $V_1$, with degeneracies 234, 108, 36, 2.
The projection of $H_t$ into these classical ground states is finite only for the first two, and the pattern of degeneracies at low energies becomes $(4,10,4,\ldots)$ for both.
This low energy pattern is the same as the one that is obtained for the CMs phase in exact diagonalization, \cite{GGNVC13} 
pointing to the fact that these two states represent the CMs phase (further evidence to support this claim is also shown below.).  
The third state is an intermediate state with the same energy as the CDW~III classical states, but which is not corrected by quantum fluctuations at any $V_1/V_2$.  
The fourth is again the CDW~I. 
The energies per site of the three phases in the $3\times3$ cluster are given by
\begin{eqnarray}
E_{\mathrm{CMs}} &=& \left\{ \begin{array}{c} \frac{V_2}{2} + \frac{V_1}{3} -0.248t \hspace{8mm} \tfrac{V_1}{V_2} < \tfrac{3}{2} \\  
\frac{2V_2}{3} + \frac{2V_1}{9} -0.055t \hspace{5mm}   \tfrac{3}{2} <\tfrac{V_1}{V_2} < 3\end{array} \right. \\
E^* &=& \frac{5V_2}{6} + \frac{V_1}{6} \hspace{22mm}   3 <\tfrac{V_1}{V_2} < 4 \\
E_{\mathrm{CDW~I}} &=& \frac{3V_2}{2}   \hspace{29mm}   4 < \tfrac{V_1}{V_2} 
\end{eqnarray}
As explained in the main text, the energy per site of the CDW~II phase $E_{\mathrm{CDW~II}}$ is lower than the classical energies of any state of the $3\times3$ cluster. 
Since the intermediate state does not have quantum corrections, its energy satisfies $E_{\mathrm{int}}>E_{\mathrm{CDW~II}}$ and it is never realized. 
The CMs state, however, has $E_{\mathrm{CMs}} < E_{\mathrm{CDW~II}}$ for small enough $V_1$ once the quantum corrections are included. 
Further evidence that this is the CMs state can be obtained by computing correlation functions. 
Fourier transforming the real space indices $ij\ldots$ and taking the combinations defined in the main text for the representations $B_2$ and $G$, we can compute any correlation function of order parameters $\left< GS \right| O_1 O_2 \ldots \left| GS \right> $ with $O = G, B_2$. 
The simplest scalars that signal the presence of charge order are simply the two point functions
\begin{align}
S_1 &= B_2^2, \\
S_2 &= |G_1|^2+|G_2|^2.
\end{align}
We have computed these correlation functions in the ground states of the $3\times3$ cluster with $V_1<3V_2$ and finite $t$ and found that they are both finite. 
However, this is not enough information to distinguish the CMs phase. 
In order to detect its particular order we have also computed the correlation functions of higher order for this ground state
\begin{align}
S_3 &= {\rm Im}(G_2^3-3G_2 G_1^2),  \\
S_4 &= B_2 {\rm Im}(G_1G_2), \\
S_5 &= [{\rm Im}(G_1G_2)]^2, \\
S_6 &= B_2{\rm Re}(G_1^3-3G_1 G_2^2),
\end{align}
and found that only $S_6$ is finite, therefore confirming the presence of the CMs phase with the strong coupling approach for $V_1<3V_2$. 
As for the CDW~III we have also compared the semicalssical and numerical structure factors and found good agreement, supporting the semiclassical interpretation of the CMs phase.

\end{document}